\documentclass[iop,apjl]{emulateapj}
\usepackage{amsmath}

\usepackage{graphicx} \usepackage{natbib}

\usepackage{subfigure} \usepackage{epsfig}

\slugcomment{{\sc Accepted to ApJ Letters:} December 18, 2012}

\begin{document}

\title{Retention of Stellar-Mass Black Holes in Globular Clusters}
  
\author{Meagan Morscher, Stefan Umbreit, Will M.\ Farr, and Frederic
  A.\ Rasio}

\affil{Center for Interdisciplinary Exploration and Research in
  Astrophysics (CIERA), and Department of Physics and Astronomy,
  Northwestern University, 2145 Sheridan Road, Evanston, IL 60208,
  USA.}

\email{m.morscher@u.northwestern.edu}
\email{s-umbreit@northwestern.edu}
\email{w-farr@northwestern.edu}
\email{rasio@northwestern.edu}

\begin{abstract}
Globular clusters should be born with significant numbers of
stellar-mass black holes (BHs). It has been thought for two decades
that very few of these BHs could be retained through the cluster
lifetime. With masses $\sim 10\, M_\odot$, BHs are $\sim$ 20 times
more massive than an average cluster star.  They segregate into the
cluster core, where they may eventually decouple from the remainder of
the cluster.  The small-$N$ core then evaporates on a short timescale.
This is the so-called Spitzer instability.  Here we present the
results of a full dynamical simulation of a globular cluster
containing many stellar-mass BHs with a realistic mass spectrum. Our
Monte Carlo simulation code includes detailed treatments of all
relevant stellar evolution and dynamical processes. Our main finding
is that old globular clusters could still contain many BHs at
present. In our simulation, we find no evidence for the Spitzer
instability. Instead, most of the BHs remain well-mixed with the rest
of the cluster, with only the innermost few tens of BHs segregating
significantly. Over the 12 Gyr evolution, fewer than half of the BHs
are dynamically ejected through strong binary interactions in the
cluster core. The presence of BHs leads to long-term heating of the
cluster, ultimately producing a core radius on the high end of the
distribution for Milky Way globular clusters (and those of other
galaxies). A crude extrapolation from our model suggests that the
BH--BH merger rate from globular clusters could be comparable to the
rate in the field.

\end{abstract}



\section{Introduction} \label{Intro}

Typical globular clusters (GCs) should form $\sim 100\, -\, 1000$ BHs
within $\sim$ 3 Myr and could retain most of them initially, if their natal 
kicks are sufficiently low (see \citealt{Wong2012} and references therein).  
With masses $\sim 10\, M_\odot$, these BHs become the most massive 
objects in the cluster after just $\sim 10\,$Myr, so their dynamics will be very
different than that of typical stars.  The presence of BHs can affect
the overall structure and evolution of the parent cluster
\citep{Mackey2008}.  BHs accreting from a stellar companion can be
visible as bright X-ray binaries (XRBs), which are in principle
detectable both in our own and other nearby galaxies
\citep{Kalogera2004}. Merging BH--BH binaries are key sources of
gravitational waves (GW) which should be detectable by upcoming
interferometers such as Advanced LIGO \citep{HarryLIGO2010}.

It is well known that the formation rate per unit mass of XRBs is
higher in clusters by orders of magnitude compared to the field (e.g.,
\citealt{Pooley2003}). This indicates that dynamics must play an
essential role in producing cluster XRBs. Prior to 2007, however,
there had not been a single detection of a \emph{black hole} XRB
inside a GC, although many had been identified in galactic
fields. This appeared to agree with many theoretical studies
suggesting that essentially all BHs within a star cluster should be
ejected through dynamical interactions on a timescale $\sim 10^9\,$yr
\citep{Kulkarni1993, Sigurdsson1993, PortegiesZwart2000, OLeary2006,
  Banerjee2010}.  Key to all these previous studies is the expectation
that BHs will segregate rapidly through dynamical friction, on a
timescale $\sim 100\,$Myr, and will succumb to the so-called Spitzer
instability \citep{Spitzer1969, Kulkarni1993}, i.e., dynamically
decouple from the cluster by forming a central subcluster consisting
primarily of BHs. The relaxation time for this small-$N$ sub-cluster
of BHs is very short, leading to prompt core collapse and evaporation.
Through dynamical interactions, some BHs will be ejected in the form
of tight BH--BH binaries that merge via GW emission within a Hubble time.

This scenario was first proposed based on simple analytic estimates by
\cite{Kulkarni1993} and \cite{Sigurdsson1993}.  The first direct
$N$-body simulations of this effect used up to $N\sim 10^4$ particles
(e.g., \citealt{PortegiesZwart2000,Merritt2004}).  Larger direct
$N$-body simulations by \cite{Banerjee2010} used $N\sim 10^5$, but
with a single black hole mass ($10\,M_\odot$) and no primordial
binaries.  Using simple dynamical models, \citet{OLeary2006} and
\citet{Sadowski2008} studied the evolution of populations of BHs that
were either completely decoupled from or in equilibrium with the
cluster, respectively (see discussion in \citealt{Downing2010a}).

Monte Carlo (MC) methods have made it possible to model realistic GCs
\emph{self-consistently} with $N\sim 10^5 - 10^6$ and significant
primordial binary fractions (e.g., \citealt{Giersz2008},
\citealt{Chatterjee2010}).  The most realistic GC models with BHs to
date are from \cite{Downing2010a, Downing2011b}, who used a
H{\'e}non-type MC code to simulate clusters with $N = 5\times 10^5$
stars, a distribution of BH masses, and primordial binaries.  These
works used analytic cross sections to determine the results of
dynamical interactions, rather than direct integration.

Previous studies suggest that most BHs are ejected on a timescale
$\sim 10^9\,$yr.  Hence, old GCs should have very few, if any, BHs
left. However, in 2007 the first candidate BH XRB \emph{inside} a GC
was detected in NGC 4472 \citep{MaccaroneNature2007}.  Since then,
several additional BH candidates have been found in clusters in other
galaxies \citep{Brassington2010, Shih2010, Barnard2011,
  Maccarone2011}.  \citet{Strader2012} discovered two stellar-mass BHs
in a \emph{Milky Way} (MW) GC (M22).  Assuming these BHs are accreting
from white dwarf (WD) companions, and using calculated formation and
survival rates from \cite{Ivanova2010}, \citet{Strader2012} estimate
that M22 has $\sim 5-100$ BHs.

Furthermore, there have been a few recent theoretical suggestions that
significant numbers of BHs could still remain in some old clusters
\citep{Mackey2008,Moody2009}.  \citet{Mackey2008} used $N$-body
simulations with BHs to explain the radius-age trend in the clusters
in the Magellanic Clouds; with different initial retention fractions
for BHs, they were able to reproduce the trend of increasing spread in
core radius with age in these systems. In some models, they retained 
up to $\simeq 100$ BHs over a Hubble time. 
 
Here we re-examine the BH retention question based on a realistic,
large-$N$, fully self-consistent MC model.  We find that that at least
some old MW clusters may indeed retain a large fraction of their
primordial BHs. This dramatically different picture for the fate of
BHs in GCs may have implications for both BH XRBs and the production of
merging BH--BH binaries.

 
\section{Method}

We use a H{\'e}non-type MC method to self-consistently model the
evolution of star clusters due to the effects of two-body relaxation,
strong binary scattering encounters, stellar collisions, single and
binary stellar evolution, and mass loss from the Galactic tidal field.
A detailed description of our code, as well as examples of its
capabilities and comparisons with other methods, can be found in
\citet{Joshi2000, Joshi2001, Fregeau2003, Fregeau2007,
  Chatterjee2010}. The code has been well tested against direct
$N$-body models whenever possible.  Since the dynamical evolution of
BHs in clusters is strongly dependent on interactions involving BH
binaries, we perform direct calculations of all strong 3-body
(binary-single) and 4-body (binary-binary) interactions using the
small-$N$ integrator {\tt Fewbody} \citep{Fregeau2004}. These interactions
are responsible for the hardening of BH--BH binaries and ejections of
BHs from the cluster. Single star and binary evolution are modeled
using the routines of SSE and BSE
\citep{Hurley2000,Hurley2002}. Orbital energy loss from GW emission is
handled within BSE for binaries retained in the cluster.  For ejected
systems, which are no longer evolved with our code, we use a
simplified timescale for GW inspiral in the weak field limit
\citep{Peters1964}.  Neutron stars and BHs receive natal kicks
with velocities drawn from a Maxwellian distribution with $\sigma$=265
km s$^{-1}$. For BHs, the kick velocity is lowered according to the
amount of material that falls back onto the final BH after the
supernova explosion, according to \cite{Belczynski2002}.

We have recently added to our code a prescription for three-body
binary formation, which is important for the evolution of BHs
\citep{Kulkarni1993, Sigurdsson1993, PortegiesZwart2000,OLeary2006,
  Banerjee2010}.  We follow a similar procedure to
\citet{Ivanova2005}, \citet{Ivanova2010}, and \citet{OLeary2006} to
obtain an expression for the binary formation rate as a function of
hardness ratio
\begin{equation}
\eta = \frac{G \, m_1 \, m_2}{r_p \, <m> \, \sigma^2}.
\label{eq:eta}
\end{equation}
We keep both the geometric and gravitational focusing contributions to
the cross-section (in contrast to \citealt{Ivanova2010}, where the
geometric part of the cross section for the third star to interact
with stars 1 and 2 is dropped).  For local number density, $n$, and
average relative velocity at infinity, $v_{\infty}$, the rate at which
two stars ($m_1$ and $m_2$) form a binary with hardness $\eta \, \geq
\, \eta_{\rm min}$ through an interaction with a third star ($m_3$) is
given by
\begin{multline}
\Gamma(\eta \geq \eta_{min}) = \sqrt{2} \pi^2 n^2
      {v_{\infty}^{-9}} \\ \times (m_1 + m_2)^5 \eta_{\rm min}^{-5.5} (1 + 2
      \eta_{\rm min}) \\ \times \left[ 1+2 \eta_{\rm min} \left( \frac{ m_1 + m_2 +
            m_3}{m_1 + m_2 } \right) \right].
\label{eq:Gamma}
\end{multline}
As we expect only dynamically hard binaries to survive
\citep{Heggie1975}, we only consider the formation of hard binaries
with $\eta \geq 5 = \eta_{\rm min}$.  We allow three-body binary
formation only for BHs.  When forming a three-body binary, we choose a
value for $\eta$ from a distribution according to the differential
rate, d$\Gamma$/d$\eta$, with lower limit $\eta_{\rm min}$. The rest
of the properties of the system are calculated from conservation of
momentum and energy.

We have checked that our MC prescription produces binaries at
a rate that is in agreement with the analytic rate from
Eq.\ (\ref{eq:Gamma}). We have also done a set of tests using a direct
N-body code \citep{Farr2007} to check that our prescription produces
hard binaries at the correct rate. Using one of our cluster snapshots,
we integrated our system of BHs for a short period of time with direct
$N$-body, and compared our binary formation probability prediction to
the actual binary formation probability in the direct integration. We
find good agreement with the direct $N$-body trials, which gives us
confidence that we are actually producing hard binaries at the correct
rate.


\section{The Evolution of a Cluster with Stellar-Mass Black Holes}

 \begin{figure}
	\plotone{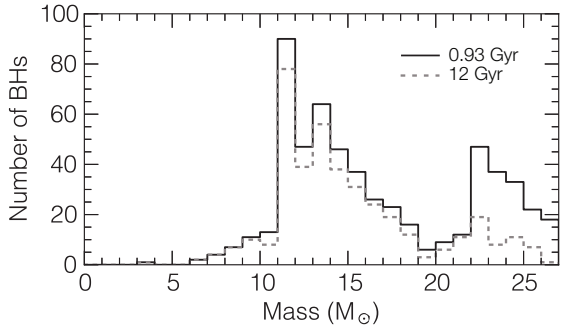}
	\caption{Mass functions for BHs present \emph{in the cluster}
          at 0.93 Gyr (solid black line) and at 12 Gyr (dotted grey
          line). The population at 0.93 Gyr is representative of the
          original population of \textit{initially retained} BHs, as
          some (about 100) are ejected from the cluster at formation
          from natal kicks. Between 0.93--12 Gyr, the most massive ($m_{bh} \geq 20\,
          M_\odot$) are ejected preferentially over the lower mass BHs
          because they have higher interaction rates.}
	\label{fig:bh_hist}

\end{figure}

We present the results of a cluster model starting with $N = 3\times
10^5$ stars following a King profile with $W_0 = 5$, half-mass radius
$r_h = 2.44$ pc, metallicity of $Z = 0.001$, and initial binary
fraction $f_b$ = 0.1.  We choose our stellar masses from the
\cite{Kroupa2001} initial mass function ranging from 0.1 --
100$\, M_\odot$. The cluster has initial total mass $M_{\rm tot} =
2.03 \times 10^5\, M_\odot$ and half-mass relaxation time $T_{\rm rh}
\approx 6.5 \times 10^8$ yr. The central escape speed 
is\footnote{Our cluster is well below the velocity dispersion 
limit of $\sim 40$ km s$^{-1}$ found in \cite{Miller2012}, above which 
growth of a massive BH through successive mergers might occur.} 
31 km s$^{-1}$. 
Only about 12\% of the BHs formed in the cluster received natal kick 
speeds above this value. We choose our remnant masses according to
\cite{Belczynski2002}, which produces BH masses in the range 
$\sim 5-30\, M_\odot$ for $Z=0.001$.
We form about 700 BHs in total, of which about 600 are retained
initially (the remainder are ejected by natal kicks). The BH mass
distribution at an early time is shown in Figure \ref{fig:bh_hist}.

\begin{figure*}
\plotone{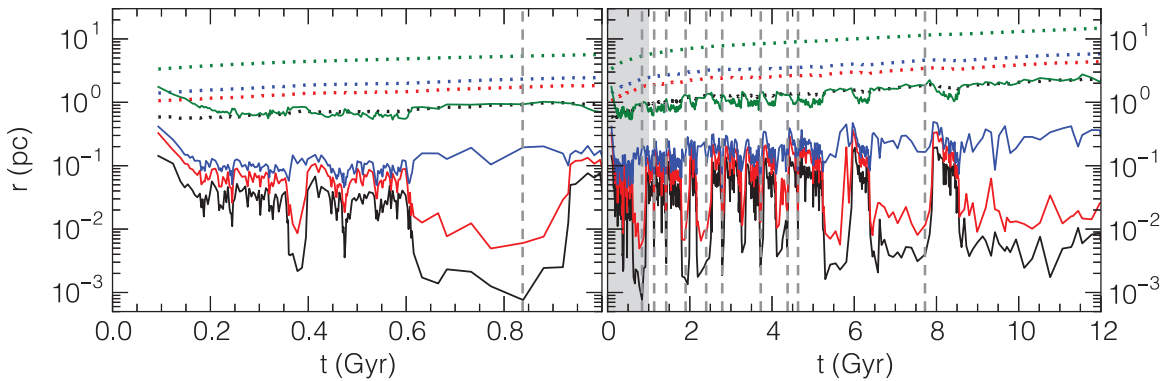}

   \caption{Evolution of the Lagrange radii of the BHs and all
     \emph{other} objects. Plots show the radii enclosing 1\%, 5\%,
     10\%, and 50\% of the mass, from bottom to top, for each category
     of object: BHs (solid curves) and non-BH (dotted curves). The left panel 
     shows the evolution over the shorter interval corresponding to the grey
     band in the right panel. The BHs segregate from the rest of the cluster 
     on a timescale of a few hundred Myr. The vertical grey dashed lines
      indicate the times when three-body binaries form. 
      Interactions involving these hard binaries cause most of the oscillations 
      of the innermost Lagrange radii. However, interactions involving 
      hard binaries \emph{not} created by three-body formation can also 
      cause the same
     effect.}
      \label{fig:lagrad}
      \end{figure*}

\begin{figure}

	\plotone{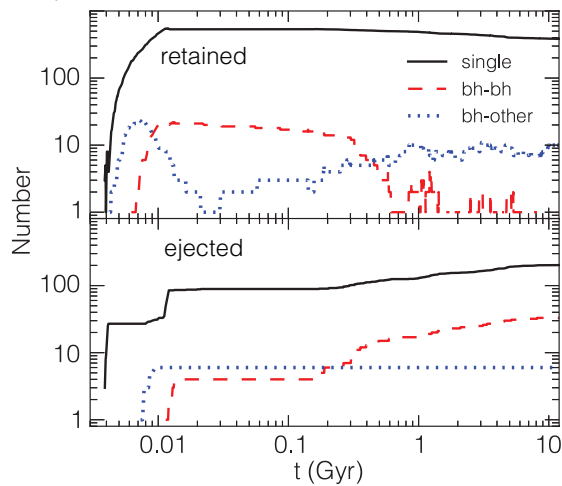}
 	\caption{Evolution of BH population retained in (top) and
          ejected from (bottom) the cluster. Each plot shows the
          number of single BHs (solid black line), BH--BH binaries
          (dashed red line), and BH-binaries with a \emph{non-BH}
          companion (dotted blue line), either retained or ejected.
          The number of BH--BH binaries within the cluster tends to
          \emph{decrease} over time until there are just a few.  Both
          single BHs and BH--BH binaries are ejected efficiently from
          about 300 Myr until about 6 Gyr, at which point the ejection
          rate slows down significantly.  This occurs in conjunction
          with a drop in the overall binary interaction rate, which is
          caused by the ongoing cluster expansion due to heating by
          the BHs. BH--other binaries tend to increase in number very
          slowly over time.  At 12 Gyr, the cluster still has 395 BHs,
          more than half of the cluster's initial population. The
          majority of the BHs are \emph{single} at all times.}
	\label{fig:bhs_time_evol}

\end{figure}

Within a few Myr, the BHs begin to segregate, leading to a central
collapse by about 400 Myr. Formation of three-body binaries and their
subsequent interactions lead to repeated core oscillations (see Figure
\ref{fig:lagrad}).  After about 300 Myr, strong binary interactions
involving the mass-segregated BH population start to become
dynamically important, and the rate of ejection of BHs (both single
and binary) increases abruptly. Ejections continue through the end of
the simulation, but the rate slows down over time. The evolution of
the numbers of single and binary BHs retained in and ejected from the
cluster is shown in Figure \ref{fig:bhs_time_evol}.  For the entire
simulation, most of the BHs are \emph{single}; in fact, beyond about
300 Myr, there are typically no more than about 10 BH-binaries in the
cluster.

Statistics of the BHs at different times are shown in Table
\ref{table:BHproperties}.  By 12 Gyr, the cluster has ejected 202
single BHs, 33 BH--BH binaries, and 6 BH binaries with non-BH
companions. Throughout the simulation, 13 BH--BH binaries merge due to
GW emission; 6 of these mergers occur within the cluster, while the
rest occur post-ejection.  Most of the BH ejections and BH--BH mergers
occur within about the first 6 Gyr of evolution. At 12 Gyr, our
model still has nearly 400 BHs, more than half of the
initially-retained population.

\begin{figure*}
\plotone{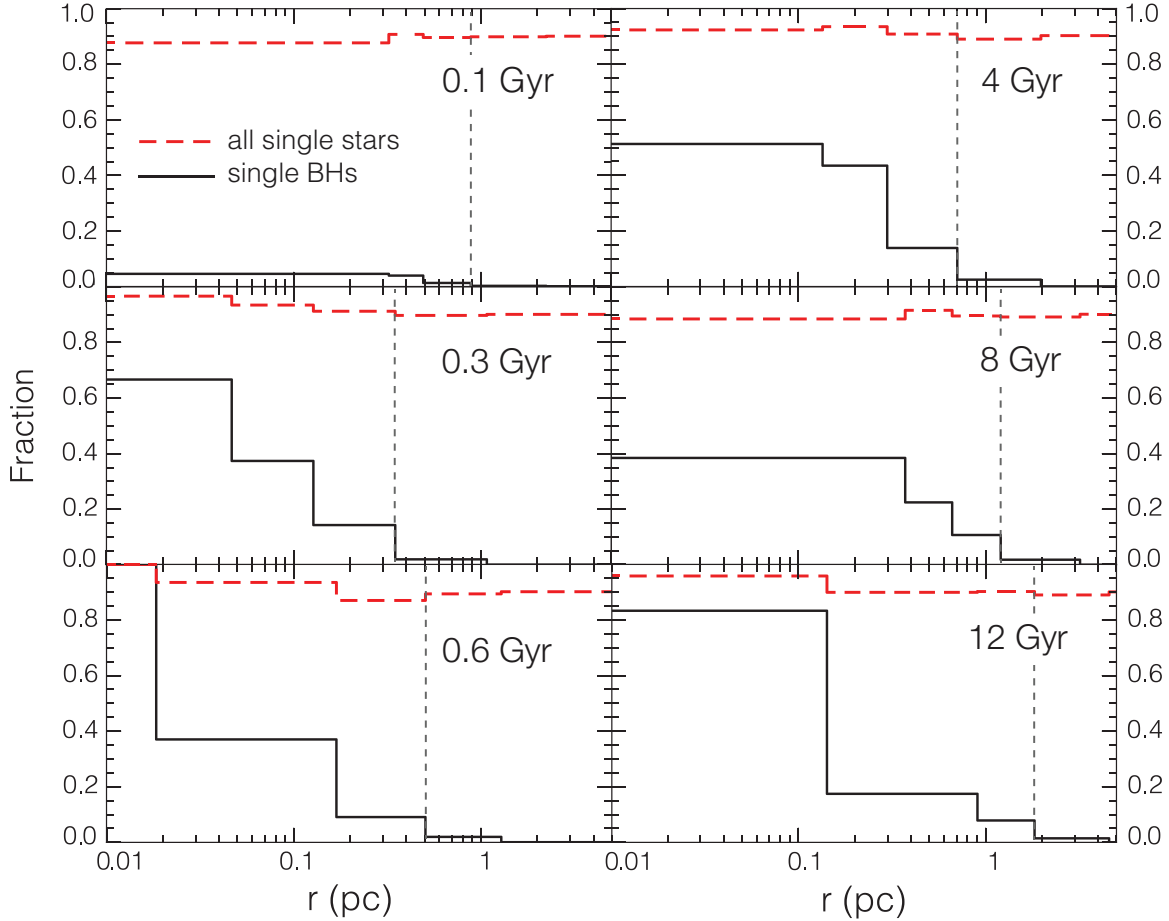}

\caption{
  The fraction of \emph{single BHs} (solid black line) and \emph{all single stars}
  (dashed red line) at several times; the remainder of the objects are binaries.  
  The innermost bin contains 20 BHs, and the number of BHs inside 
  each subsequent bin doubles (40, 80, etc.).
  The BH fraction in the central bin (containing 20 BHs) reaches unity by
  about 600 Myr (left panels), and then fluctuates between 0.4--1 for
  the rest of the simulation (right panels). Beyond the first bin, the
  BH fraction decreases, reaching negligible fractions beyond
  about 1 pc.
  The vertical grey dashed line shows the extent of the innermost 140 BHs 
  ($20+40+80=140$ contained within the first three bins),
  which is at all times less than half of the retained BH population.}
\label{fig:cum_fractions} 
\end{figure*}

In Figure \ref{fig:cum_fractions} we show the fractions of
single BHs and all single stars in radial bins at several times.  
The BH fraction in the central bin, which always contains 20 BHs,
grows to unity within about 600 Myr (left three panels), 
meaning that the innermost 20 objects are \emph{all} BHs. 
Just outside the central bin,
the BH fraction is typically less than 0.4, and it decreases to 
negligible fractions beyond about 1 pc.
 This indicates that, while the BHs do
indeed segregate to some extent, most of the BHs \emph{do not}
dynamically decouple from the cluster (i.e.\ they do not become
Spitzer unstable). All but the innermost 20 or so most massive BHs
remain well mixed with the cluster at all times.

The most massive BHs tend to be preferentially ejected from the
cluster (see Figure \ref{fig:bh_hist} and Table
\ref{table:BHproperties}). Nearly 75\% of the ejected BHs have masses
$\gtrsim 20\, M_\odot$, despite the fact that these more massive BHs
are much less common than lower mass BHs. Since the most massive BHs
sink the deepest, they tend to have the highest rates of strong
interactions, which provide the energy needed to eject them from the
cluster.

We end at 12 Gyr, a typical age for MW GCs, with the cluster having $N
= 2.47\times 10^5$ stars, $M_{\rm tot} = 1.05\times 10^5\,M_\odot$,
$r_h$ = 12.7 pc, and binary fraction $f_b$ = 0.098.  The final mass of
our cluster is just slightly larger than the median value for MW GCs
($M_{med}\approx 8 \times 10^4$; \citealt{HeggieHut2003}).  For our
model we find an observational core radius, $r_c \simeq 5-7$ pc, which
falls within the high end of the core radius distribution of the MW GC
system, and is also consistent with the range of core radii associated
with old ($\sim 10$ Gyr) GCs in the Magellanic Clouds (see Figures 1
and 2 in \citealt{Mackey2008}).

\section{Discussion and Conclusions}


The evolution and survivability of BHs in clusters, as well the effect
that BHs have on their host cluster, will depend strongly on the
degree to which the BHs are able to decouple from the cluster.  Our MC
method allows us to include realistic initial conditions as well as
all the relevant physics for studying these types of systems in
detail.  

In the most optimistic model of \citet{Mackey2008} (\texttt{run 4}),
about 50\% of their BHs ($\approx 100$) are retained over $\sim 10$
Gyr. This is slightly less than our final retention fraction (about 65
\%), but with $N$ of three times that of \citet{Mackey2008}, this
amounts to more than a factor of three difference in the actual
\emph{number} of BHs that we retain.  In contrast with
\citet{Mackey2008} who found no BH--BH mergers within a Hubble time,
we produce 13 mergers.  In clusters with low central escape
velocities, recoil kicks from strong dynamical encounters may tend to
eject BH-BH binaries before they are tight enough to merge within a
Hubble time. Although some of the Mackey et al. (2008) models do
indeed have significantly lower escape velocities than the model we
present, their \texttt{run 4} actually has a comparable escape velocity, so
this cannot reconcile the difference in merger rate.  
Instead, the discrepancy may be explained by the larger number of BHs,
 as well as the inclusion of primordial binaries, resulting in a 
 higher interaction rate in our simulation, which is consistent with the larger 
 number of ejected BHs
(but see discussion in
\cite{Downing2010a} about the competing effects of \emph{hardening}
and \emph{destruction} of BH--BH binaries, that go along with high BH
interaction rates).  We also compare to \cite{Downing2010a}, who track
BH--BH mergers in a set of MC simulations. Their model
\texttt{10low75} is most similar to ours, with the same binary
fraction and metallicity, and $N = 5 \times 10^5$, $r_{\rm h}$ = 2 pc
and $T_{\rm rh}$ = 5.25 $\times 10^8$ yr. They produced 6 $\pm$ 3
mergers (averaged over 10 simulations) within $T_H$, about half as
many as we produce in our simulation.  Agreement to within a factor of
two is reasonable, considering their use of cross sections for
predicting the outcomes of strong binary interactions (rather than
direct integration), which may overestimate the disruption rate for
tight BH--BH binaries \citep{Downing2010a}.

A crude extrapolation from our model can be used to estimate the rate
of BH--BH mergers in a Milky Way equivalent galaxy (MWEG). In our
model, the total merger rate is $\sim 1$ per Gyr.  Our Galaxy may have
had $\sim 300$ GCs (about half of which have since dissolved). We
therefore estimate a merger rate of $\sim0.3$ per MWEG per Myr from
star clusters. This is exceeds the estimated merger rate from
primordial binaries in the galactic field \citep{Abadie2010}. Thus,
our model indicates that it is important to include GCs in
calculations of the BH--BH merger rate of the Universe.

Our results indicate that at least \emph{some} old GCs could have
\emph{hundreds} of stellar-mass BHs at present. Since nearly all of
our BHs are single, our prediction is consistent with the small number
of BH XRBs detected in clusters to date. This result is timely,
considering the recent discovery of \emph{two} BH XRBs in a Milky Way
GC by \cite{Strader2012}, who suggest that there may actually be $\sim
5-100$ BHs in M22 at present. Our main conclusion is different from that 
of many other studies in the literature. This difference is not easily reconciled,
 but will be the subject of future investigations.

As has been suggested by \cite{Mackey2008}, the presence of BHs can
indeed cause heating that can lead to significant core expansion,
as we confirm with our model.
The smaller cores observed in 
MW globulars may indicate larger BH kicks than assumed in this work; 
intriguingly, \citet{Repetto2012} suggested such a change to the kick distribution 
on the basis of a population synthesis study of Galactic BHs.


\begin{deluxetable*}{c|cc|cc|cc|cc|cc}
\tablecolumns{11} \tabletypesize{\scriptsize} \tablewidth{0pc}
\tablecaption{Properties of BH population at different evolutionary
  stages: 0.93 Gyr, 3.25 Gyr, 6.5 Gyr, 9.77 Gyr and 12 Gyr. The table
  shows the number of single BHs ($N_{\rm sBH}$), number of BH-BH
  binaries ($N_{\rm BH-BH}$), number of BH--other (non-compact)
  binaries ($N_{\rm BH-other}$), number of BHs with masses above 20$\,
  M_\odot$ ($N_{\rm BH}(m\geq20\, M_\odot)$), average individual BH
  mass ($m_{\rm ave,BH}$), that are retained in/ejected from the
  cluster, at the different times.  We also show the number of BH--BH
  mergers ($N_{\rm mgr}$) that have occurred up to the time given,
  either inside the cluster or post ejection.}  \tablehead{
  \multicolumn{1}{c}{type} & \multicolumn{2}{c}{$T=0.93$ Gyr} &
  \multicolumn{2}{c}{$T=3.25$ Gyr} & \multicolumn{2}{c}{$T=6.50$ Gyr}
  & \multicolumn{2}{c}{$T=9.77$ Gyr} & \multicolumn{2}{c}{$T=12$ Gyr}
  \\ \multicolumn{1}{c}{} & \multicolumn{2}{c}{$=1.4\,T_{rh}$} &
  \multicolumn{2}{c}{$=5\,T_{rh}$} &
  \multicolumn{2}{c}{$=10\,T_{rh}$} &
  \multicolumn{2}{c}{$=15\,T_{rh}$} &
  \multicolumn{2}{c}{$=18.5\,T_{rh}$} \\ \colhead{} &
  \colhead{ret} & \colhead{ej} & \colhead{ret} & \colhead{ej} &
  \colhead{ret} & \colhead{ej} & \colhead{ret} & \colhead{ej} &
  \colhead{ret} & \colhead{ej} }

\startdata
$N_{\rm sBH}$ ...........................  & 534 & 89 & 434 & 166 &
395 & 197 & 387 & 201 & 385 & 202\\ $N_{\rm BH-BH}$
......................  & 17 & 4 & 1 & 26 & 0 & 32 & 0 & 33 & 0 &
33\\ $N_{\rm BH-other}$ ....................  & 3 & 6 & 10 & 6 & 8 & 6
& 9 & 6 & 10 & 6\\ $m_{\rm ave,BH}\, (M_\odot)$ .............. & 16.7
& 13.3 & 15.6 & 20.6 & 15.1 & 21.8 & 14.9 & 22.0 & 14.9 &
22.0\\ $N_{\rm BH}(m\geq20\, M_\odot)$ ........  & 178 & 27 & 101 &
110 & 69 & 143 & 63 & 148 & 63 & 149\\ $N_{\rm mgr}$
............................  & 4 & 0 & 5 & 1 & 6 & 6 & 6 & 7 & 6 & 7
\enddata
\label{table:BHproperties}
\end{deluxetable*}

\acknowledgements

We thank the anonymous referee for many suggestions that improved this paper.
This work was supported by NSF Grant PHY-0855592 and NASA ATP Grant
NNX09AO36G.  MM acknowledges support from an NSF GK-12 Fellowship
funded through NSF Award DGE-0948017 to Northwestern University.  The
computations in this paper were performed on Northwestern University's
HPC cluster Quest.

\newpage
\bibliographystyle{hapj} \bibliography{mybibtex}

\end{document}